\documentclass[useAMS,usenatbib]{mn2e}
\usepackage{graphicx}

\def\la{\raise.5ex\hbox{$<$}\kern-.8em\lower 1mm\hbox{$\sim$}}
\def\ma{\raise.5ex\hbox{$>$}\kern-.8em\lower 1mm\hbox{$\sim$}}

\def\msol{M$_{\odot}$ }
\def\Lsol{L$_{\odot}$ }
\def\kms{$\rm km\, s^{-1}$}
\def\cm3{$\rm cm^{-3}$}
\def\Ts{$\rm T_{*}$~}
\def\Vs{$\rm V_{s}$~}
\def\n0{$\rm n_{0}$}
\def\B0{$\rm B_{0}$}
\def\ne{$\rm n_{e}$~}

\def\Te{$\rm T_{e}$}
\def\Tgr{$\rm T_{gr}$}
\def\Td{$\rm T_{d}$}

\def\erg{$\rm erg\, cm^{-2}\, s^{-1}$}
\def\mum{$\mu$m~}

\def\L12{L$_{12\mu m}$~}
\def\F12{F$_{12\mu m}$~}
\def\agr{a$_{gr}$}
\def\Hb{H${\beta}$~}
\def\Ha{H$\alpha$~}

\def\ff{{\it ff}}

\title[The merger Seyfert galaxy NGC 6240]{Metallicity in the merger Seyfert  galaxy NGC 6240} 
\author[M. Contini]{ M. Contini 
\\
School of Physics and Astronomy, Tel Aviv University, Tel Aviv
69978, Israel \\
}

\begin{document}

\date{Accepted: Received ; in original form 2010 month day}

\pagerange{\pageref{firstpage}--\pageref{lastpage}} \pubyear{2009}

\maketitle

\label{firstpage}

\begin{abstract}

We have calculated the physical conditions throughout the merger Seyfert galaxy NGC 6240
by modelling the observed optical and infrared line ratios.
We have found that the optical spectra are emitted by clouds  photoionised by the  
power-law radiation flux from  the AGN (or AGNs), and
 heated mainly by the shock accompanying the propagation of the clouds
outwards.
The infrared line ratios are emitted from  clouds ejected from a starburst which photoionises the gas
by the black-body radiation flux corresponding to a
stellar colour temperature  of about 5 10$^4$ K. 
Both the flux from the AGN and the ionization parameters are low.
The most characteristic physical parameters  are the relatively high shock velocities ($\geq$500 \kms) and low 
preshock densities ($\sim$ 40-60 \cm3) of the gas.
The   C/H, N/H, O/H  relative abundances  are higher than solar by a factor 
$\leq$ 1.5. We suggest that those relative abundances indicate   trapping of  H into
  H$_2$ molecules rather than  high metallicities.
Adopting an initial grain radius of 1 \mum, the dust temperatures calculated in the clouds reached by the
power-law  radiation and by the black-body radiation are 81 K and 68 K, respectively.

\end{abstract}


\section{Introduction}

Colliding galaxies are  at first identified  by the morphological appearance of the encounter,
depending on the  merging evolution age, 
but they  are hardly   recognizable  through the physical properties of the new galaxy. 
Eventually, the spectra show 
the dominant role of dynamical processes such as shock waves and starbursts created   by collision,
yet the active galactic nuclei (AGNs) have a dominant role.

Evidences of galaxy collision  in the Seyfert galaxy NGC 6240  appeared
  on prints of the Palomar Sky Survey (Fosbury \& Wall 1979, hereafter FW79),
namely, a complex high luminosity structure with  substantial   
 plumes extending over 130 kpc,  large dust lanes and
chaotic appearance of the nuclear region.
The synchrotron characteristic of the radio emission, the coincidence of a central feature
seen in the optical (\Ha + [NII] emission) with radio emission, the strength of the low level ionization
lines reveal the presence of  shock waves as a result of collision.
FW79 in fact suggested   that the NGC 6240 system is the result of an encounter between galaxies.
 They  dismissed the hypothesis that the  collision  could
 trigger star formation,  comparing the observed
line ratios with  those emitted from gas ionized by   the black body radiation flux
from young stars.

On the other hand, Rieke et al. (1985) defined  NGC 6240 as a site of
exceedingly powerful bursts of star formation, involving nearly 10$^{10}$ \msol
of newly formed stars. This was suggested not only by the luminosity and 
extent of the stellar component near 2 \mum, but particularly on the basis of 
the large population of red supergiants required to account for the depth of the 
stellar CO absorption. 

NGC 6240 was  definitively  classified among mergers
by Fried \& Schulz (1983) who detected a "pronounced" double nucleus
on an S1 image tube plate in the I band and on r and I exposures taken with a
CCD camera.
The two nuclei are oriented at a position angle of 10$^o$ and separated by 1.8"
(640-700 pc, projected on the sky)
with diameters  of 2.5" and 1.5". 

Fried \& Schulz claimed that both nuclei appear to be 
non-stellar,  rejecting the hypothesis of an accidental projection of a foreground star.
The extended regions observed by  them are dominated by shock
heating on a scale of $\sim$ 6 kpc.
The binary active nucleus in NGC 6240 was confirmed  by Komossa et al (2003). Using CHANDRA,
they discovered two hard X-ray nuclei coincident with the optical-IR ones.
Each nucleus host an AGN detected in hard X-rays 
and at 5 GHz (Gallimore \& Beswick 2004).

A starburst in NGC 6240 is the source of the K-band luminosity, dominated by supergiants (Tecza et al. 2000). 
The starburst
was triggered quite recently and has  a duration time  less than 5 million yrs. 
It  fills  the central kpc, the region encompassing the two nuclei
 (Engel et al 2010). 
Heckman  et al. (1990) showed  evidence of a  starburst driven super-wind on a large scale ($>$ kpc).

At a distance of 97 Mpc (z=0.0245,  H$_o$=75 \kms Mpc$^{-1}$), NGC 6240 has an infrared luminosity  
L$_{IR}$$\sim$ 10$^{11.8}$ \Lsol (Engel et al  2010). 
It was included in the list of ultra-luminous infrared galaxies (ULIRGs) by Rigopoulou et al. (1999).
The dynamical field is characterised by
the extraordinary large stellar velocity dispersions ($\sim$ 350 \kms).
The optical spectrum of NGC 6240 shows many emission lines with
 line ratios  characteristic of a collisional dominated regime rather
than   a power-law radiation dominated flux  generally  prevailing in the narrow line region (NLR) of AGN (FW79). 

 The mid-infrared (IR) spectrum taken with the Infrared Spectrograph on the {\it Spitzer Space Telescope}
by Armus et al (2006) shows  the flux of the most significant strong fine-structure lines up to [SiII] 34.8 \mum, as well as rotational H$_2$ lines
and PAH emission features.   Armus et al  inquired  about the  buried AGN 
contribution to the bolometric luminosity.
The mid-IR spectrum was included in the survey of the AGN NLR  in the IR by
 Dasyra et al (2011).

Comparing mid-infrared emission-line properties from high-resolution Spitzer spectra of a hard X-ray (14-195 keV) 
selected sample of nearby (z $<$ 0.05) AGNs detected by the Burst Alert Telescope (BAT) 
aboard Swift, Weaver et al. (2010) found that the luminosity distribution  of the  
[O IV] 25.89 \mum, [Ne II] 12.81 \mum, [Ne III] 15.56 \mum, and [Ne V] 14.32, 24.32 \mum lines and  hard X-ray continuum  
are similar for  Seyfert 1 and Seyfert 2 populations.
 They also found  that the emission lines primarily arise in gas ionized by the AGNs. 
On the other hand, Luhman et al. (1998) 
 report measurements of the [C II] 157.74 \mum fine-structure line in ULIRGs with the Long Wavelength Spectrometer 
on the Infrared Space Observatory. They claim
 that the observation of [CII]/ FIR continuum ratio is consistent with starburst nuclei.

 In this paper we investigate the source of radiation and the physical conditions of the emitting 
gas in NGC 6240
by a detailed modelling of  the optical spectrum presented by FW79 and 
 of the mid-IR spectrum   reported by Armus et al. 

The clumpy and irregular  morphology of NGC 6240 and the relatively large FWHM of the spectral lines
indicate that, besides the photoionization flux from the AGNs, shocks  are heating and ionizing the NLR gas.
The leading role of shocks was already noticed by FW79 on the basis of the strong neutral and
low level ionization lines.
Indeed, different regions throughout  merging galaxies show  physical conditions changing on small scales
(e.g. for NGC 7212 and NGC 3393). The spectrum observed from NGC 6240 is an average of the spectra emitted from 
the different regions. Trying to disentangle  various components throughout a unique spectrum
leads to  approximated results. However, the rather peculiar line ratios observed  by FW79 suggest
that the task is worthy. We will refer to the observed spectra with the maximum precision
as we have done for  other merger galaxies (e.g. NGC 3393, Contini 2012,  NGC 7212, Contini et al 2012, 
MKN 298, Radovich et al. 2005), i.e.
trying to find some records of  collision  analysing the spectra. 
A first hint can be obtained by  modelling  the 
spectral line ratios and the spectral energy distribution (SED) of the continuum. 
Our aim, in particular, is to  distinguish  the contribution  of the
AGN from that of the  starbursts in a  strongly dominated shock wave regime. 

We will adopt composite models
which account for the flux from the active centre (AC) and/or  for  radiation from the stars,
consistently coupled with shock wave hydrodynamics.
We have found in previous investigations (e.g. Rodr\'{i}guez-Ardila et al. 2005, 
Contini et al. 2004a, Viegas et al. 1999, and references therein)
that,  not only  shocks  are important to explain high ionization level lines (e.g. [FeVII],
[FeX]) coronal lines (e.g. [SiVIII], [SiIX])  as well as low ionization and neutral lines (e.g. [OI] and [NI]), 
but also they determine the intensity and the frequency of the SED peaks in the X-ray  and IR domains.

We have run a grid of models
 in order to select the most appropriated one
reproducing the NGC 6240 observed line ratios. Building the grid
by many models we could explore which ranges of the physical parameters could
characterise  the  NLR of NGC 6240.

The  calculations of the spectra are presented in Sect. 2. 
The optical spectrum presented by FW79  and the fine-structure  
line ratios in the IR observed by Armus et al.
are investigated in Sect. 3. 
The calculated continuum SED is compared with the observational data in Sect. 4.
Discussion and concluding remarks follow in Sect. 5.

\section{The calculation code}

In this paper, the line and continuum spectra emitted by the gas downstream of the shock front 
are calculated by the
code {\sc suma}\footnote{http://wise-obs.tau.ac.il/$\sim$marcel/suma/index.htm}.
The code simulates the physical conditions in an emitting gaseous cloud under the coupled effect of 
photoionization from an external radiation source and shocks. The line and continuum emission 
from the gas are calculated consistently with dust-reprocessed radiation in a plane-parallel geometry.

\begin{figure*}
\includegraphics[width=14.cm]{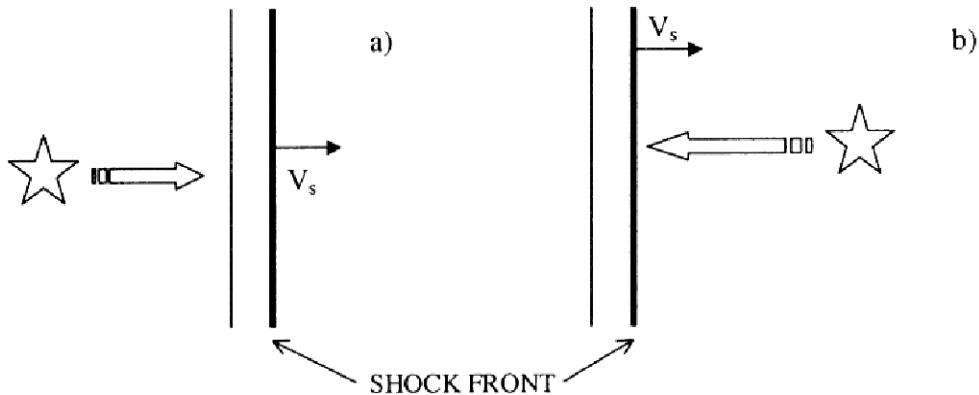}
\caption{Case a) : the cloud  withdraws from the external radiation source which is represented by the star;
 case b) the cloud approaches the radiation source}
\end{figure*}

\subsection{Input parameters}

The input parameters which refer to the shock, are represented by the 
shock velocity \Vs, the pre-shock density \n0, the pre-shock magnetic field \B0. 
The input parameter  that  refers to the radiation field  is the power-law
flux  from the active center $F$  in number of photons cm$^{-2}$ s$^{-1}$ eV$^{-1}$ at the Lyman limit,
if the photoionization source is an active nucleus, and the spectral indices  $\alpha_{UV}$=-1.5
and $\alpha_X$=-0.7.
 $F$  is combined with the ionization parameter $U$ by

$U$= ($F$/(n c ($\alpha$ -1)) (($E_H)^{-\alpha +1}$ - ($E_C)^{-\alpha +1}$)
(Contini \& Aldrovandi, 1983), where
$E_H$ is H ionization potential  and $E_C$ is the high energy cutoff,
$n$ the density, $\alpha$ the spectral index, and c the speed of light.

If  stars with a colour temperature \Ts are the photoionization source,
the number of ionizing photons cm$^{-2}$ s$^{-1}$ produced by the hot 
source is $N$ = $\int_{\nu_0}$ $B_{\nu}$/h$\nu$ d$\nu$, 
where $\nu_0$ = 3.29  10$^{15}$ Hz and B$_{\nu}$  is the Planck function. 
The flux from the star is combined with $U$ and n by $N$ (r/R)$^2$=$U$nc, where r is the radius of the hot source (the stars),
 R is the radius of the nebula (in terms of the distance from the stars), n is the density of the nebula and c is the 
speed of light. Therefore, \Ts  and $U$ compensate each other, but only in a qualitative way, because \Ts  determines 
the frequency distribution of the primary flux, while $U$ represents the number of photons per number of electrons 
reaching the nebula. The choice of \Ts and $U$   is made by the fit of the line ratios.

The secondary diffuse radiation emitted from the slabs of gas heated by the shocks is also calculated.
The  flux  from the active center or from the stars (primary radiation) and the secondary radiation are  calculated by
radiation transfer throughout the slabs downstream.

The dust-to-gas ratio ($d/g$) and the  abundances of He, C, N, O, Ne, Mg, Si, S, A, Fe relative to H,
are also accounted for. They  affect the calculation of the cooling rate.
 The dust grains are heated radiatively by  photoionization  and, collisionally, by the shock.
Changing $d/g$ will change the mutual heating and cooling of gas and dust.
The distribution of the grain radii downstream
results from sputtering.

The geometrical thickness of the emitting nebula $D$,  determines whether the model is
radiation-bound or matter-bound.

\subsection{Calculation process}

The calculations start at the shock front where the gas is compressed and thermalized adiabatically, 
reaching the maximum temperature in the immediate post-shock region (T$\sim 1.5 \times 10^5$ (\Vs/100 \kms )$^2$). 
T decreases downstream following recombination. The cooling rate is calculated in each slab. 
The downstream region is cut up into a maximum of 300 plane-parallel slabs with different geometrical 
widths calculated automatically, in order to account for the temperature gradient (Contini 2009 and references therein).

In each slab, compression is calculated by the Rankine--Hugoniot equations for the conservation of mass, 
momentum and energy throughout the shock front. Compression ($n/n_0$) downstream ranges between 4 
(the adiabatic jump) and $\geq$ 100, depending on \Vs and \B0. The stronger the magnetic field, the lower the 
compression downstream, while a higher shock velocity corresponds to a higher compression.

The ionizing radiation from an external source is characterized by its spectrum  and by 
the flux intensity. The flux is calculated at 440 energies, from a few eV to keV. Due to radiative transfer, 
the spectrum changes throughout the downstream slabs, each of them contributing to the optical depth. 
In addition to the radiation from the primary source, the effect of the diffuse radiation created by 
the gas line and continuum emission is also taken into account, using 240 energies to calculate the spectrum.

For each slab of gas, the fractional abundance of the ions of each chemical element is obtained by solving 
the ionization equations. These equations account for the ionization mechanisms 
(photoionization by the primary and diffuse radiation, and collisional ionization) and recombination 
mechanisms (radiative, dielectronic recombinations), as well as charge transfer effects. 
The ionization equations are coupled to the energy equation when collision processes dominate and 
to the thermal balance if radiative processes dominate. The latter balances the heating of the gas due 
to the primary and diffuse radiations reaching the slab with the cooling due to recombinations and 
collisional excitation of the ions followed by line emission, dust collisional ionization and thermal bremsstrahlung. 
The coupled equations are solved for each slab, providing the physical conditions necessary for calculating the slab 
optical depth, as well as its line and continuum emissions. The slab contributions are integrated throughout the cloud.

In particular, the absolute line fluxes referring to the ionization level i of element K are calculated by the 
term $n_K$(i) which represents the density of the ion X(i). We consider that $n_K$(i)=X(i)[K/H]$n_H$, where X(i) is 
the fractional abundance of the ion i calculated by the ionization equations, [K/H] is the relative 
abundance of the element K to H and $n_H$ is the density of H (by number \cm3). In models including shock, 
$n_H$ is calculated by the compression equation (Cox 1972) in each slab downstream. 
So the abundances of the elements are given relative to H as input parameters.

Dust grains are coupled to the gas across the shock front by the magnetic field. 
They are heated by radiation from the AGN and collisionally by the gas to a maximum temperature which is a 
function of the shock velocity, of the chemical composition and of the radius of the grains, 
up to the evaporation temperature ($T_{\rm dust}$ $\geq$ 1500 K). The grain radius distribution downstream is determined 
by sputtering, which depends on the shock velocity and on the density. Throughout shock fronts and downstream, 
the grains might be destroyed by sputtering.

Summarizing, the code starts by adopting an initial gas electron temperature \Te ($\sim$ 10$^4$ K) and the input parameters 
for the first slab (Sect. 2.1). Then, it calculates the density from the compression 
equation, the fractional abundances of the ions from each level for each element, line emission, 
free-free emission and free-bound emission. It re-calculates \Te by thermal balancing or the 
enthalpy equation, and calculates the optical depth of the slab and the primary and 
secondary fluxes. Finally, it adopts the parameters found in slab i as initial conditions 
for slab i+1. By integrating the contribution of the line intensities calculated in each slab, 
we obtain the absolute fluxes of each of the lines, calculated at the nebula (the same for bremsstrahlung). 
We then calculate the line ratios to a certain line (in the present case \Hb for the optical spectrum
which is a strong line), 
and we compare them with the observed line ratios.

In particular, the code accounts for the direction of the cloud motion relative to
  the external photoionizing source (Fig. 1). If the  radiation flux from the source
reaches the  shock front edge of the cloud, the switching parameter $str$=0, whereas when  the flux reaches the
edge opposite to the shock front, $str$=1.
In the former case the calculations start at the shock front and proceed until the gas is at a temperature
below 10$^3$ K (the model is radiation bound) or the calculations are interrupted when all the lines reproduce
the oberved line ratios (the model is matter bound).

In the latter case, i.e. when the cloud propagates outwards from the radiation source, the calculations 
require some iterations, until the results converge.
In the first iteration a shock dominated model is adopted ($F$=0.). We start from the shock front
in order to calculate the profile of the density  downstream. The ionization equations account only for
the secondary radiation and collisional ionization. The density as well as the secondary radiation are stored
for each of the slabs. The temperatures calculated in the first iteration are stored only in the region of
the cloud where collision processes prevail. They will be used in the following iteration.

In the second iteration the calculations start from the edge of the cloud reached by the
external radiation flux. The primary  and secondary radiation fluxes are  calculated by radiation transfer throughout
each slab  and are adopted for  the resolution of the ionization  equation system.
The secondary radiation from the photoionized side of the cloud is  stored for each slab. 

In the  following iteration all the process is recalculated
starting from the shock front using the primary flux stored in the previous iteration and the secondary
radiation from both sides of the cloud.
It is clear that the cloud geometrical thickness plays an important role.  Figs. 2 and 3 show that, if the cloud is
very thin, the cool gas region may disappear  leading to  [OI]=0 and [NI]=0.

\section{Modelling the line spectra}

We start by modelling the line spectra because they are more constraining than the continuum SED.
 In fact, the line ratios depend on the stratification of the ions downstream, that is very
specific for each model. On the other hand, the
bremsstrahlung shows the characteristic trend for gas at  temperatures $\leq$ 10$^4$ K common to all the models
at frequencies lower than 10$^{14}$ Hz.
In the  UV range it is  generally blended with the strong black body emission corresponding
to the old star population background.  Only  the maximum frequency of the bremsstrahlung peak 
depends directly on the shock velocity. 
 
\subsection{Selection of the models}

The  physical parameters are combined throughout the calculation of forbidden and permitted lines
 emitted from a shocked nebula.
The ranges of the physical conditions in the nebula are deduced, as a first guess, from the observed
lines. 

Generally, the shock velocity is determined by  both
 a close approximation  to the observed  FWHM of the line profiles
and the  best fit to the observed  line intensities ($\sim$ 400-500 \kms for NGC 6240). 
 A high \Vs increasing compression downstream, speeds up the cooling rate, leading to
enhanced low ionization level lines. 

 The [NI] 5200+ (the + indicates that the doublet is summed up) lines
constrain the model because the critical density for collisional deexcitation of the lines is very low
($<$ 2000 \cm3). Considering that compression (n/\n0) downstream is between $\sim$ 50  and $>$100 for relative high
shock velocities,  a low  \n0 ($< 100$ \cm3) is predicted.
The [OII] 3727+ and [SII] 6717+ lines  can also  be drastically reduced by collisional deexcitation at high electron densities
(\ne $>$ 3000 \cm3).

Adopting a composite model (shock + photoionization)  some interesting issues should be explained.
We fit the spectral lines   adopting an \n0 which is combined with the
other input parameters, including the preshock magnetic field \B0.
The magnetic field  has an important role in models accounting for the shock.
The stronger \B0 the lower the compression. So, lower densities are compensated 
by a  lower magnetic field. 
We have  adopted   \B0  = 10$^{-4}$ G, which is suitable to the NLR of AGN (Beck 2011).

In pure photoionization models, the density $n$ is constant throughout the nebula, while in a shock wave 
regime, $n$ is calculated downstream by the compression equation in each of the single slabs. 
Compression depends on $n$, the magnetic field $B$ and the shock velocity \Vs.
In models accounting for the shocks, both the electron temperature  $T_{\rm e}$  and the electron density \ne 
are far from  constant throughout each cloud, 
showing the characteristic profiles in the downstream region (see Figs. 2 and 3).
In particular, throughout each cloud,  the density reaches its upper limit  downstream and  remains nearly constant,
while the electron density decreases following recombination.
A high density, increasing the cooling rate,  speeds up the recombination process
of the gas,  enhancing  neutral lines.
In fact, each line is emitted from a region  of gas  at a different \ne and \Te,
depending on the ionization level and the  atomic parameters characteristic of the ion.
So, even  sophisticated calculations reproduce approximately the highly inhomogeneous
conditions of the gas,  leading to some discrepancies between  calculated and observed line ratios.

The radiation intensity of the flux from the AGN is evaluated from the [OIII]/[OII] line ratios.
The  higher $F$, the higher
 [OIII]/[OII]. Moreover,
a high $F$ maintains the gas ionized far from the source  yielding enhanced  [OI]
and [SII] lines.  Recall that these lines behave similarly because S first ionization
potential (10.36 eV) is lower than that of O (13.61 eV).

The  geometrical thickness ($D$) of the emitting clouds is mainly determined by [OI]/\Hb
because [OI] is emitted from gas at relatively low temperature,
at  the cloud edge opposite to the shock front in the inflow case, or in the centre of the cloud
for the  outflow case (see Figs. 2 and 3).
The geometrical thickness of a cloud   establishes whether a model is  radiation-bound or matter-bound.
A first analysis of the data in NGC 6240 shows that the models are mostly radiation- bound.

The parameters are intermingled  therefore the effect of a single
 parameter on the final spectrum is not always straightforward,
particularly for  outflow models in which the low and intermediate ionization level lines 
come from both sides of the cloud (Figs. 2 and 3).

During the modelling process, if a satisfactory fit is not found for all the lines,
a new iteration is started with a different set of input parameters. When a line ratio is not reproduced,
we check how it depends
on the physical parameters and we decide consequently how to change them.
Each line ratio has a different weight.  We generally
consider that the observed spectrum is satisfactorily fitted by a model when the strongest line ratios are
reproduced by the calculation within a 20\%
discrepancy and weak line ratios within 50\%.

 The final gap between  observed and  calculated line ratios  is  due to the  observational errors  both random
and systematic, as well as to the uncertainties of the atomic parameters  adopted by the code,
such as recombination coefficients, collision strengths etc., which are continuously updated,
and to the choice of the model.
The final model is constrained by the SED of the continuum.

Summarizing, the parameters within a certain range  derive from the line ratios. Then they are
refined phenomenologically by the detailed modelling of  the spectrum.
The set of the input parameters which leads to the  best fit of the observed
line ratios  determines the physical and chemical  properties of the emitting gas. These are
the "results"  of our calculations. 

\subsection{Comparison of calculated with observed line ratios}

We will  consider first  the optical spectrum   which shows oxygen lines from three ionization levels        
 constraining the choice of the models.

\subsubsection{The optical lines}

The spectrum observed by FW79 is  reported in Table 1 (top panel, second column).
It was taken with the Anglo-Australian Telescope at 3.6m ESO.
The absorption line redshift corresponds to a Hubble distance of 145 Mpc,
adopting H$_o$= 50  \kms/Mpc (FW79). The line FWHM corresponds to
700$\pm$100 \kms after correction for instrumental broadening.
The nuclear region of the galaxy suffers substantial extinction
The equivalent width of \Hb emission  after correction for Av=4 mag,
corresponds to an observed flux above the atmosphere of 1.7 10$^{-14}$ \erg. The
isotropic luminosity in the rest frame of the source is 1.8 10$^{42}$ erg s$^{-1}$.

The spectrum  shows many emission lines  from low ionization species.
 The [OII]/[OIII]  $>$ 6 line ratios ([OII]/\Hb=12.3 and [OIII]/\Hb=1.6) are unusually high
in AGN, even  for a shock dominated regime.

In the  FW79  observed spectrum \Ha and [NII]6548+6584 lines are summed up, whereas
the calculated \Ha and the [NII] 6548+6584 line ratios to \Hb
 are given separately. In fact,  [NII] 6548+6584/\Hb is  important to determine the N/H
relative abundance. We can predict that  in the observed corrected spectrum 
\Ha/\Hb $\sim$ 3  for gas  in the NLR physical conditions.
 In fact, in reddening corrected spectra emitted by gas at relatively low densities 
($\sim$ 10$^3$ -- 10$^4$ \cm3) \Ha/\Hb ranges between 3.10  for  \Te= 5000 K and 2.86 for \Te= 10,000 K (Osterbrock 1989).
Unfortunately in  NGC 6240 the [SII]6716, 6731 doublet which indicates the density of the emission gas,
is not resolved.

In Table 1 we compare  model results with the optical spectrum observed and corrected by  FW97
in the top panel. The optical line ratios refer to \Hb=1. The absolute flux of \Hb calculated 
by the models is given in the last row of the top panel. 
In the bottom section of Table 1, we present the input parameters adopted
by the models, followed by the relative abundances C/H, N/H and O/H which show some
variance from the solar ones  (C/H=3.3 10$^{-4}$, N/H=9.1 10$^{-5}$, O/H=6.6 10$^{-4}$, Ne/H=10$^{-4}$,
Mg/H=2.6 10$^{-5}$, S/H=1.6 10$^{-5}$, Fe/H=3.2 10$^{-5}$ Allen 1976).

\begin{table*}
\begin{center}
\caption{Comparison of calculated with observed spectra}
\begin{tabular}{llllllllll} \hline  \hline
   line                &  obs$^1$ &m1$_{pl}$ &m0$_{pl}$ & m2$_{pl}$ &m0$_{sb}$ & m1$_{sb}$ &m2$_{sb}$ $^2$      & m$_{AV}$ $^3$ \\ \hline
\ [OII] 3727+          & 12.3      & 12.       & 11.    &11.2       &10.       & 9.      & 9.3     & 10.       \\
\ [NeIII] 3869+        &  0.31     &  0.79     & 0.92   &0.76       & 0.06     & 0.45    & 0.4     & 0.475      \\
\ H${\gamma}$          & 0.34      & 0.44      & 0.45   &0.45      & 0.46     & 0.45    &  0.45    & 0.45     \\
\ [OIII] 4363          &  $<$0.25  & 0.32      & 0.38   &0.18       & 0.026    & 0.24    & 0.2     & 0.2 2     \\
\ \Hb                  &  1.       & 1.        &  1.    &1.         & 1.       & 1.      & 1.      & 1.        \\
\ [OIII] 5007+         &  1.8      & 4.28      & 5.5    &2.5        & 2.       & 3.      & 2.6     & 2.6        \\
\ [NI] 5200            & 0.43      & 0.27      & 0.38   &1.8        & 0.002    & 0.08    & 0.2     & 0.54     \\
\ [NII] 5755           & $<$ 0.10  & 0.096     & 0.08   &0.07       & 0.04     & 0.07    & 0.07    & 0.07     \\
\ [OI] 6300+           & 1.41      & 1.2       & 1.47   &3.         & 0.006    & 0.4     & 0.9     & 1.2      \\
\ \Ha+[NII]6584+       & 8.25      & 3.3+5.12  & 3.1+5.5&3.1+5.3    &3.+5.     & 3.2+3.7 & 3.2 +5. & 3+5.0    \\  
\ [SII] 6716+          & 2.62      & 1. + 1.2  & 1.6+1.7&1.8+1.7   &0.2+0.23  &0.74+0.93& 0.24+0.32& 1.2      \\ 
\ \Hb (\erg)           &1.7e-14    & 1.18e-3   & 1.1e-3 &1.15e-3   &8.8e-3    & 1.3e-3   & 1.9e-3   &   -        \\ \hline
\ [ArII] 6.98          & 0.3       &0.20       & 0.22   & 0.25      & 0.035    & 0.18    &0.26     &0.26       \\          
\  [SIV] 10.5          & 0.02    &0.04       & 0.04      & 0.046  & 0.026    & 0.035     &0.007    &0.014      \\
\  [NeII] 12.8         &  1.       & 1.        & 1.     &  1.       &  1.      &   1.    & 1.      &  1.        \\
\  [NeV]14.3           & 0.048    & 0.046     & 0.05   & 0.038     & 0.036    & 0.049     & 0.04   & 0.04      \\
\  [NeIII]15.5          &0.32       & 1.95      & 2.22   & 2.7       & 0.29     & 0.39    & 0.4     & 0.8          \\
\ [SIII] 18.7          &0.103      & 1.83      &0.93    & 1.6       & 1.7      & 0.73    & 0.3     &  0.7      \\
\ [NeV] 24.3           & $<$ 0.02  & 0.06      & 0.07   & 0.05      & 6.3e-3   & 0.07    & 0.05    &  0.05     \\
\  [OIV]25.9            & 0.2       & 0.33      & 0.35   & 0.28      & 0.23     & 0.31    & 0.25    & 0.25          \\
\ [FeII] 26            &  0.12     & 2.18      & 1.9    & 2.06      & 0.08     & 1.28    & 0.19    &  0.5        \\
\ [SIII] 33.48         & 0.14      & 3.14      & 1.1    & 2.1       & 1.47     & 0.81    & 0.5     &  0.8       \\
\ [SiII] 34.8          & 1.4       & 7.5       & 6.2    & 6.0       & 0.42     & 3.      & 3.6     &  4.        \\
\ [NeII] 12.8 (\erg)   &193.1e-14  & 1.04e-3   & 1.e-3  & 7.2e-4    &5.6e-3    &9.6e-4   & 1.4e-3  &    -        \\ \hline
\ [CII]158 (\erg)      &2.57e-12   &1.18e-3    & 1.2e-3 &2.9e-4     &2.8e-4    &1.7e-4   & 8.3e-4  &    -        \\  \hline 
\  \Vs(\kms)           & -         & 900       & 900    & 500       &  300     & 500     & 500     &   -         \\
\  \n0 (\cm3)          & -         & 40        & 40     & 40        &  60      & 60      & 68      &   -         \\
\  $F$ $^4$            & -         & 3.e8      & 3.e8   &3.e8       & -        & -       & -       &    -        \\
\   $U$                & -         & -         &  -     &-          & 0.06     & 0.002   & 0.002   &    -        \\
\  \Ts (K)             & -         & -         & -      &-          & 3.e4     & 5.e4    & 5.e4    &    -         \\
\  $D$ (cm)            & -         & 6.e17     & 1.23e17&1.e17      & 9.e16    & 1.e19   & 1.e19   &    -           \\
\ $str$                & -         & 1         &   0     & 1        & 0        & 1       & 1       &    -          \\ 
\ C/H                  & -         & 4.3e-4    & 3.3e-4  & 4.3e-4   &4.3e-4    &  4.3e-4 & 4.3e-4  &    -          \\
\ N/H                  & -         & 1.5e-4    & 1.5e-4  &1.5e-4    &1.5e-4    & 1.5e-4  & 1.5e-4  &    -           \\
\ O/H                  & -         & 9.6e-4    &  8.6e-4 & 8.6e-4   &9.6e-4    & 9.6e-4  & 9.6e-4  &    -          \\ \hline

\end{tabular}

\end{center}

\flushleft
 $^1$ data in the optical  come from Fosbury \& Wall (1979),  in the mid-IR  from Armus et al (2006).
 [CII] 158  comes from Luhman et al (1998); the absolute flux of \Hb, [NeII] 12.8 and [CII] 158 are observed
at Earth but the models are calculated at the nebula.

 $^2$ calculated by Si/H=2.e-5, S/H=4.e-6, Ar/H=8.3e-6 and Fe/H=3.e-6.

 $^3$ calculated by a relative weight of 0.3 for model  m2$_{pl}$ and of 0.7 for model m2$_{sb}$.  

 $^4$ in number of photons cm$^{-2}$ s$^{-1}$ eV$^{-1}$ at the Lyman limit
\end{table*}

We try to reproduce the [OIII]5007+/H$\beta$ line ratio which is rather weak in NGC 6240,
 by readjusting in particular $F$ and \Vs.
Then, we consider  [OII]/\Hb, which depends on \Vs and \n0.
The  [OII]3727+ doublet is  very strong
compared with the spectra emitted from the NLR of AGNs.
This suggests that
 the flux from the active center  is  low (log $F$ $\leq$ 9). Therefore,
a shock dominated  regime is predicted, which is characterised by  relatively
high [OII]/[OIII] ($\geq$ 1).
This does not mean that photoionization is negligible,  in fact the  active nuclei 
have been  observed.
Relatively high    [OI]/\Hb  shows that the models are  radiation- bound.

The models which reproduce approximatively the data were selected among  hundreds in a large grid.
It is expected that  the  large number of  the code input parameters  may induce degeneracy,
namely different combinations of the parameters could lead to many models reproducing the data
within the accepted ranges (within 20\% for strong lines and 50 \% for weaker lines).
As a matter of fact, a degeneracy can arise, for example, from the density and the magnetic field which are 
directly correlated, but in the present modelling we use  \B0=10$^{-4}$ gauss for all models.
Moreover,  the line intensities I$_{\lambda}$  result from integration 
(I$_{\lambda}$=${\Sigma}$ ${\Delta}$I$_{\lambda}$) on the cloud slabs downstream which  reach a maximum
number of 300 and vary from model to model. The ${\Delta}$I$_{\lambda}$ for each ion is calculated 
adopting  density,  temperature and
 flux  calculated in the specific slab. 
So the probability of finding the same spectrum for different sets of input parameters
is very low. We give  in the Appendix a  selection of  models, 
which give a hint about the role of the different parameters.
We refer to the oxygen line ratios [OIII]5007+/\Hb, [OII]3727+/\Hb and [OI]6300+/\Hb.
The grid in Table A1 shows that only two models reproduce these line ratios  by 9\% , 28\% and 53\%  (model m2$_{pl}$)
and by 2\%, 58\% and 15\% (model m1$_{pl}$).  Models m1$_{pl}$ and  m2$_{pl}$
are reported in Table 1. 
They were calculated adopting  outflow of the cloud from the galaxy centre.

We show in Table 1 also a model corresponding to a relatively high velocity (m0$_{pl}$), selected from the 
grid (Table A1)  which represents the case
of inflow. This case is less probable  considering  the starburst  super-wind on a large scale predicted by
Heckman et al. (1990). Moreover, the calculated [OIII]5007+/\Hb   disagrees from the  observed line ratio
 by $\sim$ 70\%.

\begin{figure*}
\includegraphics[width=8.8cm]{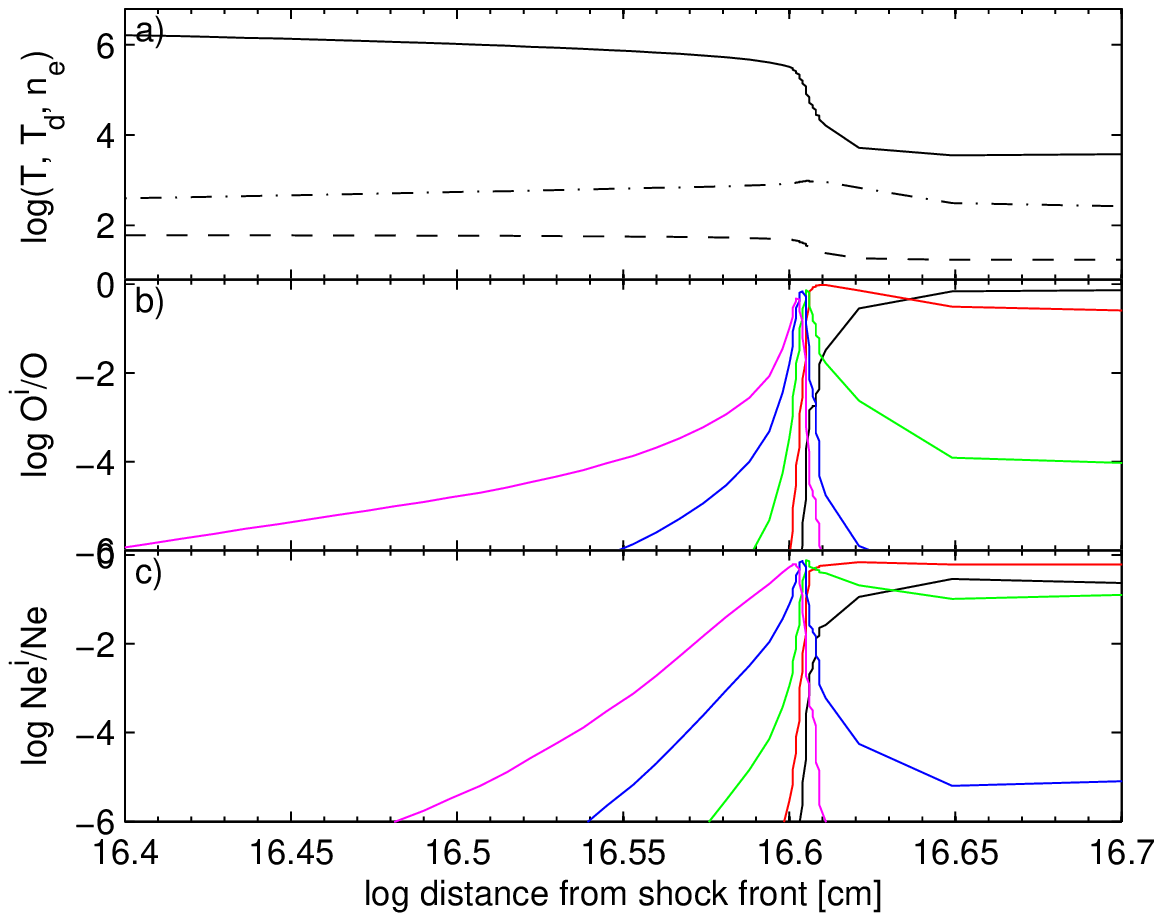} 
\includegraphics[width=8.8cm]{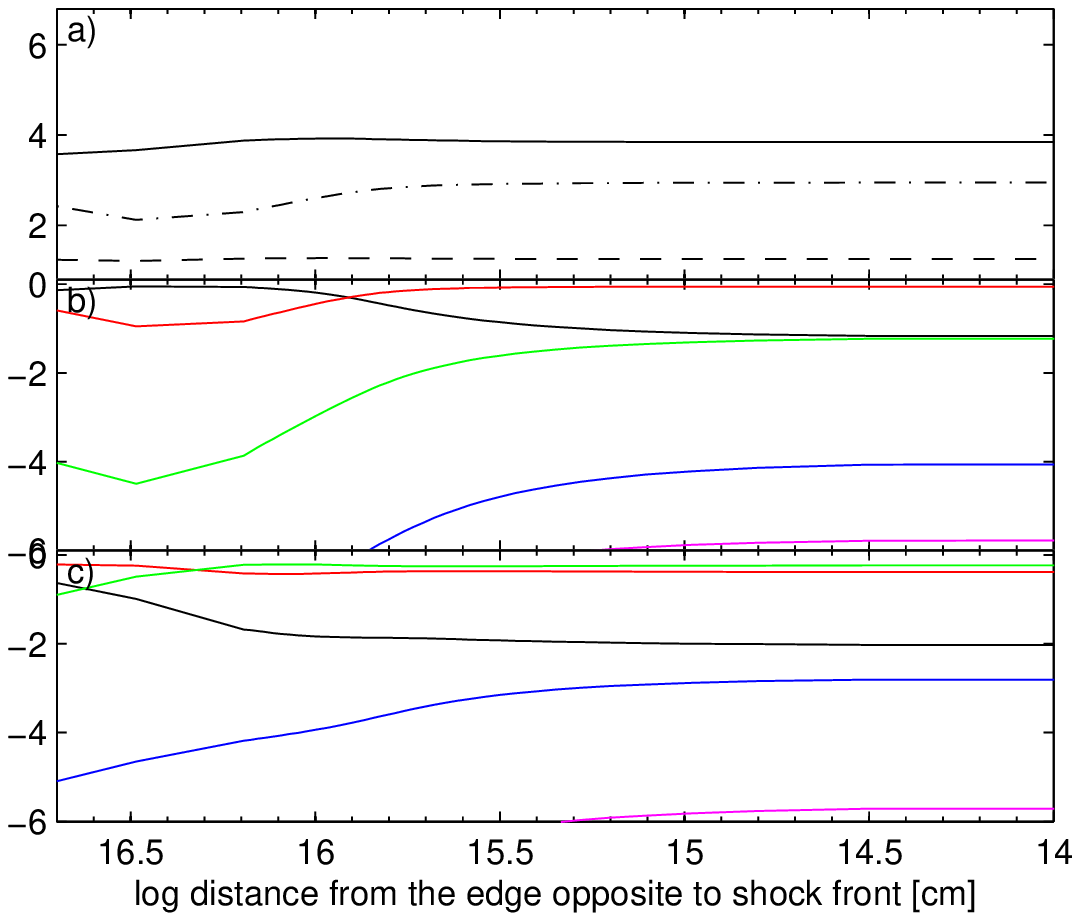} 
\caption{The physical conditions throughout a cloud corresponding
to model m2$_{pl}$ (Table 1). The shock front is on the left of the left diagram; the
cloud edge reached by the flux from the AGN is on the right of the right diagram (see text).
Top panel : T (solid line) , \Td (dashed line), \ne (dot-dashed line) .
Middle panel :  The fractional abundance of the oxygen ions : black : O$^0$/O; red : O$^+$/O; 
green O$^{++}$/O; blue : O$^{3+}$/O; magenta : O$^{4+}$/O.
Bottom panel : the same as the middle panel for Ne ions.}

\includegraphics[width=8.8cm]{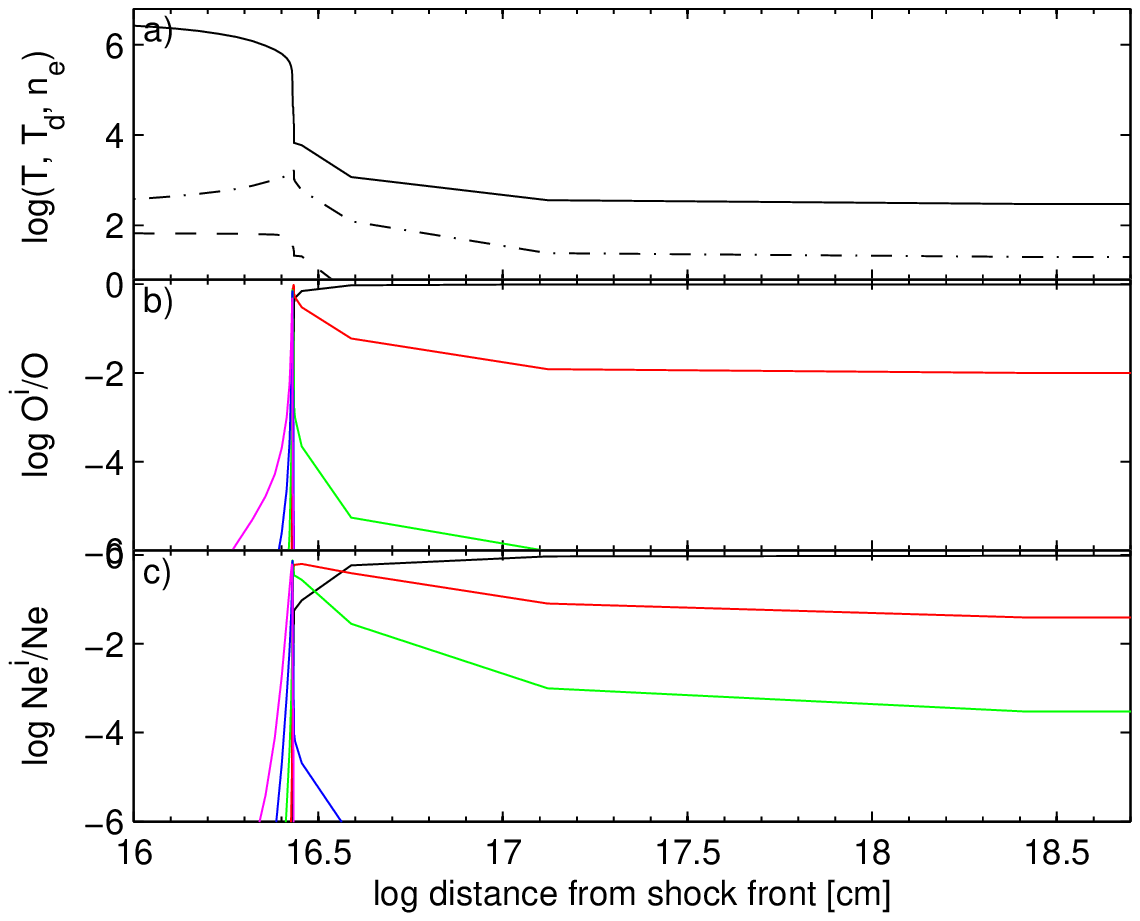} 
\includegraphics[width=8.8cm]{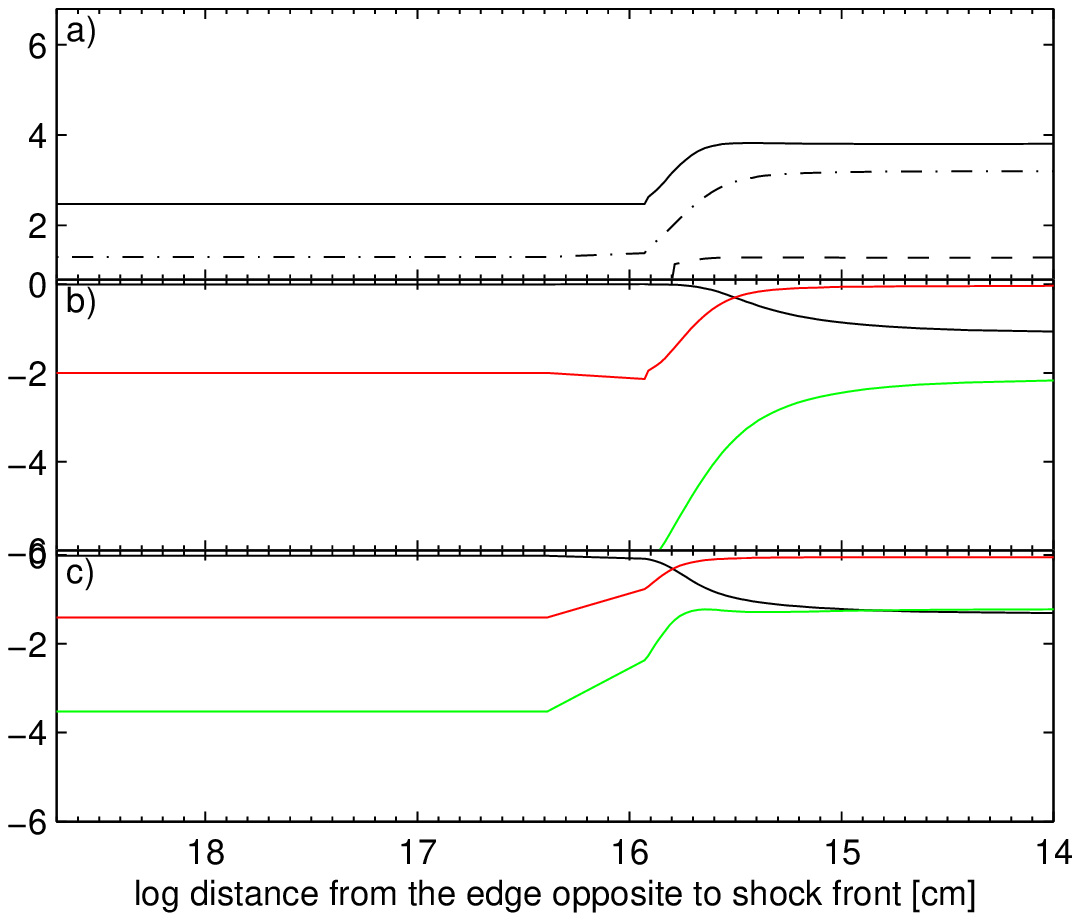} 
\caption{The same as for Fig 2 for  a cloud corresponding
to model m1$_{sb}$ (Table 1).}
\label{fig0}
\end{figure*}

Models m1$_{pl}$ and m2$_{pl}$ represent clouds ionized and heated by the power-law  (pl) flux from the AC.
They both correspond to a relatively low flux  ($F$=3 10$^8$ in 
number of photons  cm$^{-2}$ s$^{-1}$ eV$^{-1}$ at the Lyman limit), a relatively low preshock density \n0=40 \cm3
and  represent clouds propagating outward from the AC, although
 with different  velocities (\Vs= 900 and 500 \kms, respectively). 
Recall that shocks are also accounted for in  radiation
dominated models.

So far we have discussed the spectra emitted from clouds photoionised by the AGN.
However, the starburst in NGC 6240 is also a strong  photoionization source.

Models  m1$_{sb}$ and m0$_{sb}$, represent 
  outflow from the starburst and inflow, respectively. 
 The m2$_{sb}$ model represents also outflow in the starburst case. We will address it
regarding the IR lines.
The inflow case  must be dropped because
underpredicts by some orders of magnitude the [OI]/\Hb line ratios,
 as was predicted by FW79.
 The model representing  gas ejected from the starburst, m1$_{sb}$ is more appropriated  to the physical
picture, but it   reproduces the [OII]/\Hb, [OIII]/\Hb 
and [OI]/\Hb line ratios by 27 \%, 55 \% and 72 \%, respectively.  
Although models m0$_{sb}$ and m1$_{sb}$ refer to the   starburst (+shocks), we  adopted
the same relative abundances as those adopted in the pl radiation dominated models 
in order to obtain a rough fit  to the observed [OII] and [OIII] line ratios to \Hb.

We could not find any shock dominated model ($F$=0, $U$=0)  which could  consistently reproduce the
[OII]/\Hb and [OIII]/\Hb line ratios    as can be noticed from Table A1. 
These results indicate that in  NGC 6240 the shock contribution is always combined with that of radiation.

To better understand  the line ratios calculated by the models, we present in  Figs.  2 and 3
the distribution of the  temperature T (in K), electron density \ne (in \cm3) and grain temperature
\Td (in K) in the top panels. The fractional abundance of oxygen and neon
ions throughout a cloud which propagates outward from the AGN are shown in the middle and bottom panels,
respectively. 
In Fig. 2 the radiation flux reaching the  cloud is a power-law characteristic of AGNs. 
In Fig. 3 the cloud is ejected from the starburst. The radiation flux is a  black-body.
The clouds in both Figs. 2 and 3 are divided into  two halves represented by the left and right diagrams. 
The shock front is on the left of the left diagram,
while the radiation flux reaches the opposite edge (right edge of the right diagram). 
The left diagram represents the region downstream
  close to the shock front and the X-axis scale  representing distance from shock front,
is  logarithmic.  The right diagrams  show the conditions  downstream far from the shock front, 
close to the edge illuminated by the radiation flux.
The distance from the  illuminated edge  is given by a reverse logarithmic X-axis scale. 
 Thus the conditions  throughout a cloud  are seen with critical resolutions at the two edges.

The calculations show that the
 maximum temperatures in the immediate post-shock region downstream (Figs. 2 and 3, top panel)
 result 1.2 10$^7$ K and 3.75 10$^6$ K
for \Vs=900 \kms and \Vs = 500 \kms, respectively, leading to X-ray emission (Fig. 4).
The cooling rate  downstream  determines
 the distribution of the electron densities and temperatures, and consequently,
the intensity of the emission lines and of the continuum.

In composite models
for outflowing clouds (Figs 2 and 3, middle panel)), [OIII]/[OII] line ratios are higher from the photoionised side
of the cloud because radiation heats the gas at temperatures of 2-3 10$^4$ K which are suitable to the [OIII] line
ionization potential. Nevertheless,  for NGC 6240 [OIII]/[OII] $<$ 1, because $F$ is relatively low.
The [OII]/[OIII] and [OI]/[OIII] ratios are higher
 from the shock front edge because the temperature drop characteristic of shock dominated
models leads to a large region of gas at T $\leq 10^4$ K, adapted to strong [OII] and [OI] lines.
The two sides are bridged by secondary radiation which
maintains the gas at  about 10$^4$ K. So in NGC 6240  the shock  dominates,  but
photoionization cannot be neglected, in particular because the AGN are observed.

Modelling the line ratios, we had to exclude  preshock densities higher than 40-60 \cm3 and $d/g$ ratios 
higher   than 4 10$^{-3}$
 because  they  did not  improve the fitting of the line spectrum and continuum SED (Table A1). 
 The cooling rate by line and continuum emission of the gas is $\propto$ n$^2$ 
and the cooling rate by collisionally heating of dust grains is $\propto$ $d/g$ (Dwek 1987).
 Higher densities and higher $d/g$  speeding up the cooling rate downstream by a constant factor
will change the distribution
of the fractional ions (Contini 2004b). The emitted spectrum will change accordingly.

Moreover, oxygen and carbon  are   strong coolants (we cannot refer here to He whose lines  were not observed), 
 namely,  the cooling rate depends   also on the O/H and C/H relative abundances.
Adopting  C/H solar or slightly higher than solar (3.3-4.3 10$^{-4}$) and O/H  higher 
than solar 
(8.6 10$^{-4}$-9.6 10$^{-4}$) the calculation results were  acceptable. 
 In fact  higher relative abundances increase the cooling rate only in the  zones downstream where the gas reaches the
conditions adapted to  emit  strong lines,  i.e. not constantly  throughout the
whole downstream region. 

Notice that O/H values are higher than solar  referring to 
 Asplund et al. (2009),  Allen (1976), and Anders,  Grevesse  (1989),  who give
 O/H = 4.9 - 6.6 - 8.5 10$^{-4}$, respectively.
 Relatively high [NII] 6548+/\Hb ratios, fitting the data, were calculated by models adopting N/H = 1.5 solar.
The [NeIII]/\Hb line ratio is overpredicted by the models m1$_{pl}$ and  m2$_{pl}$
because the calculations account for both the ${\lambda}$3869 and ${\lambda}$3940 lines, while the observations 
refer only to the ${\lambda}$ 3869 line.
Finally,
relatively high [OI]/\Hb and [NI]/\Hb line ratios result generally  from  pl radiation-bound dominated models 
(see Contini 2004a).

\subsubsection{The infrared lines}

 In the   second panel of Table 1,  the infrared line ratios reported by  Armus et al (2006) are shown.
They  presented flux and  FWHM of fine-structure narrow lines observed from NGC 6240.
The NGC 6240 line fluxes    reported by Dasyra et al (2011) who collected  mid-IR lines  of
 type 2 Seyfert galaxies for  [SIV]10.5, [NeII]12.8, [NeV]14.3, [OIV]25.9, [NeIII]15.5  and [OIV]25.9
from the {\it Spitzer} cathalog   correspond to those of Armus et al. 
The lines are generally the strongest in the IR range
(see Figs. 2 and 3). They refer to three ionization levels of Ne, constraining the models.

We  show in Table 1  the observed line ratios to [NeII] 12.8,  
which are independent from the   neon relative abundance, at least  two of them,
and we compare them with the results of models which were selected  by the fit of the optical spectrum.
 The [NeII]  absolute fluxes are also shown
in the last row of the Table 1 second panel, both  observations at Earth (col. 2) and models calculated at the nebula
(col. 3-8).

 We first refer to the lines reported in the survey of  Dasyra et al.
The  data  are reproduced by 2 - 42 \% by the starburst dominated model (m1$_{sb}$) which appears in column 7
of Table 1. 
 The spectra dominated by a power-law radiation flux 
overpredict by a factor $>$ 6 the [NeIII]15.5/[NeII]12.8 line ratio. 
Our results show that, on the contrary,  the starburst model (m1$_{sb}$) 
 reproduces  the infrared  line ratios  within  43 \%.
Recall that although the [OIII]5007/\Hb and [OII]/\Hb line ratios were  reproduced 
within the observational error by the starburst models,
the  neutral optical lines  were underpredicted by  factors of 3 up to $>$10, suggesting  that even if  
starbursts are present,
the neutral lines emitted by the surrounding clouds  are too weak to be  observed.

 If we refer only to the IR lines, both inflow and outflow models are valid, however,
if we consider also their contribution to the optical lines, indeed the inflow model 
should be dropped because  underpredicting the [NeIII] 3869/\Hb and the
[OI]6300/\Hb line ratios by factors $>$ 50. A  more  physical constraint consists in
 the starburst super -wind indicated by Heckman et al. 
Moreover, the blue wings of [NeV] and [OIV] seems  redshifted relative to [NeII] and [NeIII]
profiles (Dasyra et al 2011).

We consider now the S, Si, Ar and Fe lines presented by Armus et al.
The [ArII]/\Hb ratio is reproduced by Ar/H twice solar. Ar is not trapped into dust grains
because of its atomic structure, while Si, S and Fe can be included in grains and molecules.
We have run model m2$_{sb}$ representing outflow from the starburst and depleted elements.
The input parameters were slightly modified in order to save the good  fit  to the optical
line ratios. The results which are given in column 8 of Table 1, show that the 
IR  [SIII] lines are  overpredicted by the model by a factor $> $2, while the optical
[SII] lines are underpredicted by a factor of 4.6 adopting S/H=4. 10$^{-6}$ (0.25 solar).
This indicates that the optical and IR sulphur lines come from different regions
and that  in the IR line emitting  region  sulphur is trapped deeply into molecular clouds.

 In the third panel of Table 1 the [CII] absolute flux presented by Luhman et al (1998) is reported.
The absolute fluxes are calculated  by the models adopted to calculate the other IR lines.
 Luhman et al claim that [CII] fluxes are very low  for NGC 6240 (and for Arp 220) and that 
NGC 6240 shows a [CII]/ FIR continuum ratio consistent with starburst nuclei.

 In the bottom panel of Table 1 the input parameters are shown.
The high \Vs adopted  by the models to fit the observed FWHM of the line profiles,
  confirm that shocks have a leading role in NGC 6240.
The  starburst dominated models  indicate
that a relatively low ionization parameter $U$, diluted by distance, is consistent with a large
geometrical thickness  $D$ of the emitting cloud.

Summarizing, the results of modelling presented in Table 1 show that 
the  AGNs dominate the  optical spectrum, while the starbursts better explain the
IR lines. The emitting nebulae are most probably located in different regions.

In the last column of Table 1 we show the averaged line ratios obtained by summing up model
m2$_{pl}$ and model m2$_{sb}$ with  relative  weights of 0.3 and 0.7 respectively.
The fit of some line ratios to the data improve, but not for all the lines . In fact,
different lines are emitted from different regions of the galaxy, so a real picture should 
account for many different conditions.

\section{The continuum SED}

\begin{figure*}
\includegraphics[width=13.1cm]{fg4a.eps}
\includegraphics[width=14.cm]{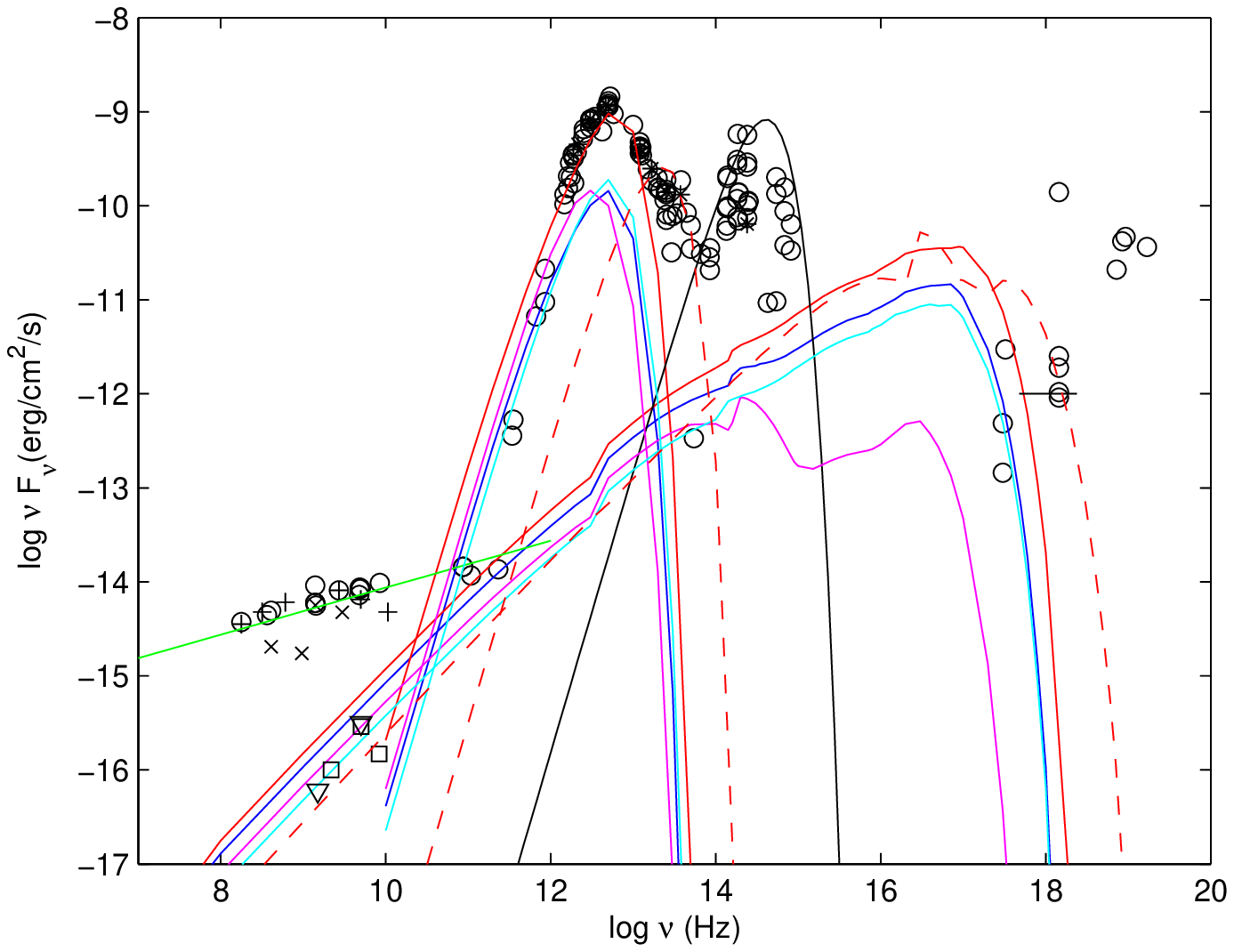}
\caption{The SED of the continuum. The data  are  explained in the text.
Blue lines : model m2$_{pl}$; red lines : m1$_{pl}$; magenta lines : m0$_{sb}$;
cyan lines :  m1$_{sb}$; red dashed lines : model calculated by \Vs=2000 \kms and \n0=500 \cm3;
green line : synchrotron radiation; black line : black body flux from background stars.
}
\label{fig3}
\end{figure*}

NGC 6240 is a luminous infrared galaxy.
This can be noticed through the continuum SED which
is  composed by the very  rich set of observations, covering the  
frequency range from radio to X-rays (Fig. 4, top panel).

The data  from the NED 
(Large et al. 1981, Condon et al. 2002, 
White, Becker 1992, Condon et al. 1996, Wright, Otrupcek  1990,
  Bennett et al.  1986, Griffith  et al. 1995, Laine  et al. 2006, Nakanishi et al.  2005, Iono et al.  2007,
   Wilson et al. 2008, Yang, Phillips 2007, Benford  1999, Spinoglio et al. 2002, Negishi et al.  2008
Brauher et al.  2008, Moshir et al 1990,  Lahuis et al. 2007, Armus et al. 2006, Egami et al. 2006,
 Golombek  et al. 1988, Sanders et al.  2003, Siebenmorgen et al. 2004, Klaas et al. 2001, Forster et al. 2004,
  Veilleux et al.  2009, Marshall et al.  2007, Allen  1976, 2MASS.  2003, Rudy et al.  1982,
 Spinoglio et al. 1995, De Vaucouleurs, Longo 1988, De Vaucouleurs et al. 1991, Gonzalez-Martin et al.  2009, 
Brinkmann et al. 1994,
 Lutz et al.  2004, Gonzalez-Martin et al.  2006, Beckmann et al. 2009, Sazonov et al. 2007, Gower et al.  1967) 
 are represented in the bottom panel of Fig. 4 by open circles.

Even if the NED data are taken with different apertures and at different times, 
Fig. 4 shows that they   very consistently
follow the trend of a power-law flux (synchrotron radiation) in the radio frequency range, a  composite black-body
infrared emission from dust in the IR,  the black body radiation  in the optical-near-UV
which bounds  the  fluxes of the old star population. 
Moreover, they   indicate X-ray  emission in the soft and hard energy ranges.

We have added the data in the radio observed by Hagiwara et al (2011)  from the northern source N2, which   shows the strongest
fluxes at epochs 2003-October-November (open triangles) and 2009-June (open squares). Hagiwara et al used the European Very Long Baseline Interferometry Network.
The + and  x refer to the low resolution observations and the flux densities  seen in the interferometer
observations by the Cambridge One-Mile Telescope (COMT) and the Jordrell Bank Interferometers (JBI) and 
Cambridge 5-km telescope (FW79 
and references therein), respectively. The errorbars  of the latter  are large,  
particularly for the datum at 10$^4$ MHz.
We have omitted the errorbars from  all the data in all the ranges far sake of clarity
 in the bottom diagram of Fig. 4 in which we compare the data with model calculations.
We have also added the  {Spitzer} IR data presented by Sargsyan et al. (2011) (collected from
the $AKARI$ catalog, from the $IRAS$ Faint Source Catalog and from the 2Mass Catalog) which are indicated by the asterisks.
The models which were  selected by the fit to the line spectra are used to   reproduce the continuum SED.

The data in the radio  from the NED
agree with  COMT observations (FW79). They follow the  trend of the
calculated synchrotron radiation flux emitted by the Fermi mechanism at the shock front (Bell 1978).
However the JBI data (FW79) show a slightly steeper trend and lower fluxes intensities.
The trend of the Hagiwara (2011) data is definitively different,  following the bremsstrahlung 
calculated by the models presented in Table 1. Moreover, the Hagiwara data  are lower by a factor of $>$ 10.

The most reasonable explanation  is that the data from the NED and from COMT refer to a large area and
embrace the clouds shocked by the starburst super-wind  (Heckman et al 1990).  
 The data from the NED are broad band observations, mostly integrated from maps.
Not only radio - optical- UV  radiation will  be emitted from the clouds,
 but being  carried by the super-wind, 
 radio synchrotron radiation  will be emitted at the shock fronts and X-rays  at the  shocked edges.
As the starburst fills the whole region between the AGNs and beyond , the same is valid for clouds
photoionised by the pl radiation.
On the other hand,  Hagiwara observations  are focused towards a well defined region. So most probably,
we see only the bremsstrahlung from the photoionised edge of the clouds.

The far IR fluxes  come from different regions in different
conditions. Notice, for example, that in the millimetric at the same frequency  ($\sim$ 3 10$^{11}$ Hz) 
the synchrotron radio tail and the reradiation by dust appear from different observations.
Moreover, the data between $\sim$ 10$^{14}$ and $>$ 10$^{15}$ Hz are nested within the Planck
function representing the old star (T=4000 K) background.

Tezca et al  (2000)   wonder about  the dust heating cause  in the IR,
namely, by the AGN or by the starburst.
In our models clouds  reached by radiation from the AGNs and from  the starburst are also
heated and ionized by the shocks. 

 As we have explained in  sect. 3, 
the continuum from the models shows the characteristic bremsstrahlung slope at frequencies
lower than 10$^{14}$ Hz common   to all the models, because emitted from cool gas (T$<$ 10$^3$ K).
The slopes between 10$^{14}$ and 10$^{15}$ Hz  depend on the flux from the AGN or from the
stars which heat the gas to $\sim$ 3-4 10$^4$ K, changing from model to model. 
At higher frequencies the slope is dictated by the shock
which can heat the gas at high temperatures depending on \Vs (sect. 2).
 Therefore we have included in Fig. 4 (bottom panel)  model m0$_{sb}$ which is calculated by \Vs=300 \kms,
for comparison.

A higher \n0 will enhance the bremsstrahlung which is $\propto n^2$. In Fig. 4 we refer in particular to the SED of the continuum.
The models, which are calculated from the emitting gas, are shifted on the Y-axis in order to reproduce the observed data
at Earth. The shift is proportional to r$^2$/d$^2$,
where r is the distance
of the gas nebula from the  AC and d is the distance of the galaxy from Earth.
So we find r=5 kpc for the clouds ionized by the AGN radiation  downstream of a shock with \Vs=500 \kms, r=5.4 kpc 
for the clouds ionized by the AGN and \Vs=900 kms and
r=3 kpc for the clouds ionized by the starburst.

 We have run a  shock dominated ($F$=0) model with \Vs=2000 \kms and \n0=500 \cm3, \agr=0.5 \mum, $d/g$=0.002 by mass.
The calculated SED nicely fits the data in the soft-X-ray range and completes the modelling of the IR reprocessed emission.
The maximum temperature reached by the grains is \Tgr=288. K.  These high velocity clouds would be
located at r=97 pc from the galaxy centre.
This model contributes only to high ionization lines ([FeVII], [FeX], etc.). It contributes by less than 10\%
to the spectra calculated by the  starburst model and  less then 1\%  to the pl calculated lines
presented in Table 1.

 Dust reprocessed radiation in the IR is calculated consistently with gas emission
by the mutual heating and cooling of grains and gas atoms. 
Collisional  processes dominate
at the high temperatures which refer to the relatively high shock velocities in NGC 6240.
Therefore,  dust grains
would be heated collisionally by the gas to relatively high temperature
at the cloud shock front edge, but here  the grains
 are easily sputtered for shock velocities   \Vs $>$ 200 \kms.
Adopting  \agr=1 \mum in the present models,  the grains are reduced by sputtering
to $\sim$ 0.5 \mum.
The maximum temperature of dust  for model m1$_{pl}$ is T$_d$=81 K and for model m1$_{sb}$ is
T$_d$=68 K.
The models were calculated adopting $d/g$= 4 10$^{-3}$ by mass.

 Armus et al 
in their analysis of the continuum SED, refer to three dust components at 27.1 $\pm$ 0.3 K (cold),
at 81.4 $\pm$ 1.8 K (warm) and at 680 $\pm$ 12 K (hot) in order to reproduce the IR dust reprocessed bump.
Our results agree with the warm temperature, but the hot temperature is much lower.
The cold temperature is included in the calculations reproducing the warm one.
Indeed. the model adopted by Armus et al about  grain properties and location  is different from our
which accounts for collisional heating of the grains. Moreover, dust emission is calculated consistently
by models that are constrained by the spectra emitted from the gas.

Komossa et al claim that a large part of the X-ray is extended.
Soft X-rays are emitted downstream of shocks with the characteristic velocities
(500-900 \kms)  that were calculated by modelling the line spectra. 
The shocks   accompany the outward  motion of the clouds carried by the super-wind.
The starburst embraces a large region, therefore the soft X-ray emission is spread
throughout the galaxy.

CHANDRA data are integrated between 2-10 keV (Fig. 4, top panel).
Although  soft-X-ray data are explained by bremsstrahlung  emitted by
the high velocity gas,  hard X-rays  are  hardly reproduced by the  bremsstrahlung from  gas
in the conditions suitable to explain the optical-IR line spectra.
In fact X-ray observations at 2-10 keV reveal the presence of iron 
line emission at 6.4 keV from both nuclei, which is most prominent in the
southern nucleus and classifies this galaxy as a binary AGN (Hagiwara et al 2011, Komossa et al 2003).
The strong Fe K$\alpha$ line is not produced in a starburst superwind but originates in
cold material illuminated by a hard continuum spectrum.

\section{Discussion and concluding remarks}

The Seyfert galaxy NGC 6240 is the result of  collision and  merging.
Merging  objects  generally include 1) a double active nucleus,
2) stars and starbursts in the central region and in the edges, even beyond them, and 3) shocked
fragmented clouds. Therefore, the  models adopted for the calculation of the spectra   account for the power-law 
radiation flux from the AGNs,  black-body radiation flux from the starburst and shock waves.

We have calculated the physical conditions throughout the  galaxy
by modelling the optical, infrared line ratios and  the continuum SED.
The best fitting models were selected from a large grid.

\subsection{The physical picture}

Modelling suggests that  the clouds photoionised by both the AGN  flux and by the starburst  move outwards,
i.e. they are not gravitationally falling towards the black holes. The starburst region embraces
the zone between the AGNs and beyond it and a starburst super-wind carries the clouds outwards (Heckman 1990).

We have found that   the flux from  the AGN (or AGNs)
$F$ is  similar to the lowest  ones found
in LINERs (e.g. Contini 1997) and LLAGNs (e.g. Contini 2004a).
It leads to  optical line ratios   best fitting the data.
The pl radiation flux photoionises the clouds
 and heats the gas at maximum temperatures of 3-4 10$^4$ K. 
The gas is  mainly heated by the shock to maximum temperatures of 3.75 10$^6$ K and 1.2 10$^7$ K  
accompanying  the clouds outwards with shock velocities \Vs $\sim$ 500-900 \kms, respectively. 
 A high \Vs combined with a low \n0  ($\sim$ 40 \kms) leads to the high [OII]/[OIII] observed in NGC 6240.

 A distance r $\sim$ 370 pc from the   AC is calculated for the  gas reached by the AGN radiation
and emitting the observed optical line ratios,
 combining the \Hb absolute flux calculated at the nebula
by   model m2$_{pl}$ (Table 1) with \Hb observed at Earth :
H${\beta}_{abs}$ r$^2$= H${\beta}_{obs}$ d$^2$, where d is the distance from Earth and H${\beta}_{obs}$=1.7 10$^{-14}$ \erg (FW79).
The distance r  is a lower limit.
A higher distance should result accounting for the filling factor,
\ff=10$^{-3}$ - 10$^{-4}$ (Heckman et al 1990). The results obtained for r by modelling the continuum in Sect. 4
lead to \ff= 0.074.

The infrared spectrum reveals the starburst. We have found that the  IR line ratios are 
emitted from gas photoionised by a 
starburst corresponding to a stellar colour temperature \Ts  $\sim$ 5 10$^4$ K.
The gas is included in   geometrically thick  clouds  ($D$= 3 pc). 
The ionization parameter  is low ($U$=0.002), indicating that the 
 starburst region is extended and   the photon flux is  diluted on its way to the emitting gas.

In Fig. 5 we plot  [NeV]/[NeII] versus [NeIII]/[NeII] for both the Type I and Type II
AGN samples of Dasyra et al (2011). We chosed only Ne lines in order to avoid the relative
abundance problem. Fig. 5 shows  that the two samples almost coincide, suggesting
that the AGN is not the main source of ionization. Indeed the starbursts and the
shocked gas throughout the NLR dominate, as  also appears by comparing
 NGC 6240  with the AGN samples  in Fig. 5.

\begin{figure*}
\includegraphics[width=14.cm]{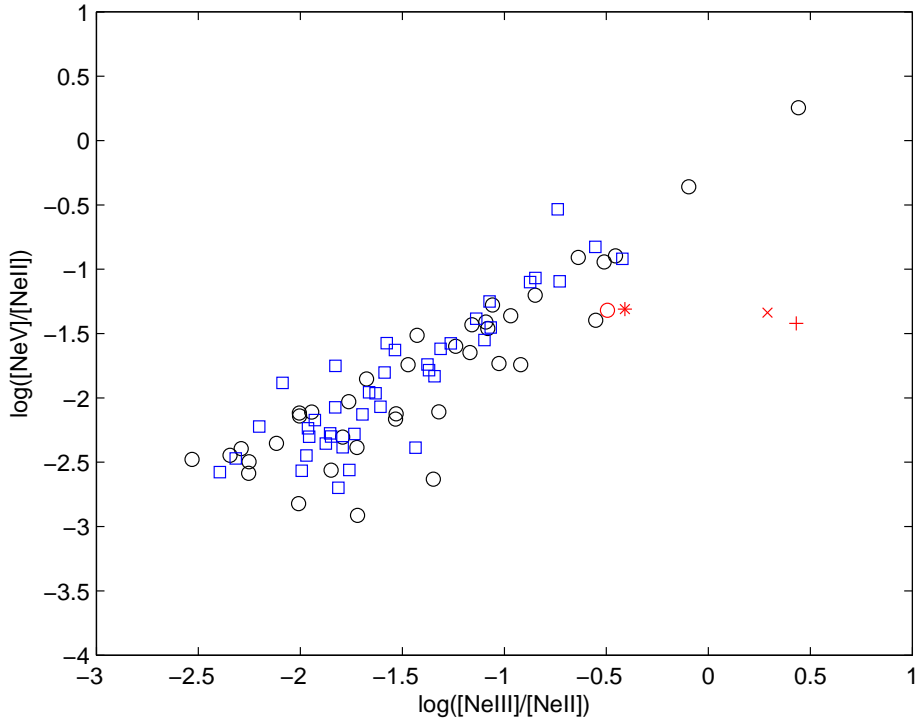}
\caption{[NeV]/[NeII] versus [NeIII]/[NeII] for the sample of Type 1 and Type 2 AGN  
presented by Dasyra et al. (2011) : blue squares and black circles, respectively. 
The red circle represents NGC 6240; the red asterisk
indicates the starburst model result. Red  + and red X represent the AGN models
calculated by \Vs=500 \kms and 900 \kms, respectively.
} 
\label{fig3}
\end{figure*}

The  clouds emitting the [NeII] 12.8 lines are at a distance r $\geq$ 4.3 kpc from the starburst, 
which is calculated by combining
the [NeII] absolute flux calculated by model m1$_{sb}$ (Table 1)
with [NeII] observed  at Earth  (Dasyra et al 2011) :
[NeII]$_{abs}$ r$^2$= [NeII]$_{obs}$ d$^2$, where d is the distance of NGC 6240 to Earth.
 A distance of 4.3 kpc was found comparing the calculated to the observed continua,
leading to \ff $\leq$ 1.

 The models presented in Table 1 are constrained enough to allow to calculate the location of the
[CII] emitting region. Combining the flux  observed at Earth with that calculated at the nebula,
 we obtain r= 9 kpc  for the
 AGN (by model m2$_{pl}$)
and 11 kpc for the starburst (by model m2$_{sb}$), adopting a filling factor of 1.

We found in agreement with FW97 that the 
preshock densities are relatively low ($<$ 100 \cm3) compared to those calculated in the NLR of Seyfert galaxies
($>$ 100 \cm3), but  higher than  those of the ISM ($>$ 1 \cm3). 
The preshock densities are  relatively low  in  both the clouds ionised by  the starburst and by the AGN.
The characteristic low density in the NLR of NGC 6240  suggests that  the gas from the ISM 
was inflated inside the colliding
galaxies  and mixed up, diluting their former densities. 
Alternatively, both the colliding galaxies were characterised by  low density gas
similar to  that  found near the centre of the Milky Way.

The shock velocities are relatively high  (\Vs=500-900 \kms) and similar for  most of the clouds which are carried 
by the wind,  confirming that the high shock velocities arised from an  extended collision event.
They are slightly higher  than the observed FWHM of the lines ($\sim$ 400 \kms),
indicating that shocks   are stronger than the bulk velocity of the clouds in some positions.

The  similarity of some of the physical conditions predicted by the  AGN and  starburst models 
(e.g. low densities, high velocities and  relative abundances slightly higher than solar) is
peculiar, indicating that  the clouds are  most probably  remnants of shells ejected  from the starburst.
The matter swept up  by the blast wave accelerates outward and fragments via Rayleigh-Taylor instabilities.
Relatively  small geometrical thickness of the clouds ($D$ = 0.03 - 0.2 pc) are calculated in the neighbourhood of the AGNs.
However, they are large enough  to include a large zone of neutral gas (Fig. 3). Here, the electron temperature and 
densities are kept relatively high by the secondary radiation, so the neutral emission lines ([NI] and [OI]) 
are high enough to fit the observed [NI]/\Hb and [OI]/\Hb line ratios.
The clouds which better reproduce the infrared line spectrum are ionised by
the starburst and have  geometrical thickness ($D$= 3 pc) higher by a factor $\geq$ 10. 
Neutral line ratios are lower than in the clouds close to the AGNs,  due to
a large  region of cold gas inside the cloud.
Different geometrical thickness are in agreement with an underlying turbulence (Contini \& Goldman 2011)

\subsection{Relative abundances of the  heavy elements in the gaseous phase}

The influence of merger-induced inflows, enrichment and gas consumption,  galactic winds, etc. 
is  important in determining the nuclear metallicity. The central metallicity is primarily a competition between the inflow 
of low-metallicity gas and enrichment from star formation (Torrey et al 2012). 

We have  obtained the best fit to the spectra by  C/H= 1. - 1.3 solar, N/H=1.5 solar,  O/H= 1.45 solar
and Ar is twice solar.
The relative abundances derived from  our modelling of the NGC 6240 emission spectra are not common in mergers at low z.
 Before considering  enrichment for instance from star formation, let us  consider  some  evidences about
grains and molecules in a collision dominated regime such as that in NGC 6240. 

Tacconi et al (1999) report the result of the CO J=2$\rightarrow$1 line observations. Half of the CO flux is concentrated
in a thick disc structure situated between the two radio/infrared nuclei, with a diameter $\leq$ 500 pc.
Strong CO bands were already found in NGC 6240 by Rieke et al (1985) indicating giant and supergiants,
where dust is formed.
Also H$_2$ v=1$\rightarrow$0 emission is centered between the radio/infrared nuclei (Joseph et al 1984).

 Mac Law \& Glover (2011) and Glover \& Mac Law, (2012) calculations
 on the formation of H$_2$ and CO in
clouds, show  that the CO molecules are easily dissociated for low densities ($\sim$ 30 \cm3) while H$_2$
molecules are still  present,  partially removing H from the gaseous phase. 
Strong molecular hydrogen emission has been detected
by Rieke et al (1985) and by Becklin et al (1984) in NGC 6240.

The 1-0S(1)H$_2$ emission line is one of the most powerful found in
any galaxy to date (Joseph et al 1984), probably excited by shocks (Tecza et al 2000, etc.),
while CO abundance around the nuclei is very likely to have been significantly
reduced by X-ray irradiation from the AGN.
 It seems that  the H$_2$ emission is excited purely by thermal mechanisms   (Sugai et al 1997)
  and H$_2$ lines are emitted between the double nuclei. 
From its excitation mechanism and its peak position,   the emission is from a global shock 
caused by a galaxy-galaxy collision. 

We suggest that   abundances of heavy elements  relative to H are high in NGC 6240 due to trapping of hydrogen 
into H$_2$ molecules, removing  it from the gaseous phase.
High  C/H, N/H,   O/H and Ar/H are therefore an indirect record of  collision,
They do not indicate  high metallicities, but they reveal the formation of
 H$_2$ molecules in a large collision event. 
We cannot predict the  true C/H, N/H and O/H relative abundances,
perhaps lower than solar.
Yet, the remnants of the  inflated gas can explain Heckman et al discovery of
 two very bright knots located about 8.5 kpc west and 2.5 kpc WSW
of the nucleus. They have line widths less than 100 \kms FWHM and \Ha is
about 15 times brighter than [SII] 6717+ and at least 25 times brighter than
[NII] 6584. Heckman et al. claim that these knots are probably giant HII 
regions with \Ha uncorrected luminosities of
about 4  10$^{39}$ and 9  10$^{39}$ erg s$^{-1}$, respectively and
 that  [NII] and [SII] lines suggest very low metal abundances,
 less than 20\% solar. 
We suggest that they represent the original inflowed matter, not processed by the super-wind. 

On the other hand, elements which are easily trapped into grains and molecules are strongly depleted from the gaseous phase.
Fe is generally included into small grains and therefore easily sputtered. The depletion of Fe indicates  that 
 depleted matter was included
at collision of the galaxies. S and Si can be included into molecules, but downstream of shocks at \Vs=500-900 \kms
everything is evaporated. So they were also  originally depleted in the included matter, in agreement with
Heckman et al.

\section*{Aknowledgements}
I am grateful to the referee for many interesting comments which improved the presentation
of the paper.
This research has made use of the NASA Astrophysics Data System (ADS) and the NED, which is operated
by the Jet Propulsion Laboratory, California Institute of Technology, under contract with NASA.

\section*{References}

\def\ref{\par\noindent\hangindent 20pt}

\ref Allen, D.A.  1976, ApJ,  207,  367
\ref Anders E.,  Grevesse N. 1989, Geochimica et Cosmochimica Acta 53, 197
\ref Armus, L. et al. 2006, ApJ, 640, 204
\ref Asplund M., Grevesse N., Sauval A.J.,  Scott P. 2009, ARAA, 47, 481
\ref  Beck, R. 2011, arXiv:1112.1823
\ref Becklin, E.E., DePoy, D., Wynn-Williams, C.G. 1984, Paper presented at the
Infrared Detector Workshop, Laramie, Wyoming, May 15-16, 1984.
\ref Beckmann, V. et al. 2009, A\&A, 505, 417
\ref Bell, A.R. 1978, MNRAS, 182, 147
\ref Benford, D.J.  1999 vol. Thesis p. 1
\ref  Bennett,C.L., Lawrence, C.R., Burke, B.F., Hewitt, J.N., Mahoney, J.
     1986, ApJS,  61, 1 
\ref Brauher, J. R.; Dale, D. A.; Helou, G.  2008, ApJS, 178, 280
\ref Brinkmann, W.,   Siebert,  J.,  Boller, Th. 1994, A\&A, 281, 355
\ref Condon, J.J., Helou, G., Sanders, D.B.,  Soifer, B.T. 1996, ApJS, 103, 81
\ref  Condon, J.J.,Cotton, W.D., Broderick, J.J. 2002, AJ, 124, 675
\ref Contini, M. 2012, arXiv:1206.3060
\ref Contini, M., Cracco, V., Ciroi, S. 2012, A\&A, submitted
\ref Contini, M., Goldman, I. 2011, MNRAS, 411, 792
\ref Contini, M. 2009, MNRAS, 399, 1175
\ref Contini, M. 2004a, MNRAS, 354, 675
\ref Contini, M. 2004b, A\&A, 422, 591
\ref Contini, M., Viegas, S.M. 2001 ApJS, 132, 211
\ref Contini, M.; Viegas, S. M.; Prieto, M. A.	2004, MNRAS, 348, 1065	
\ref Contini, M. 1997, A\&A, 323, 71
\ref Contini, M., Aldrovandi, S.M. 1983, A\&A, 127, 15
\ref Cox, D.P. 1972, ApJ, 178, 169 
\ref Dasyra, K.M. et al. 2011, ApJ, 740, 94 
\ref De Vaucouleurs, A., Longo, G. 1988, 
    Catalogue of visual and infrared photometry of galaxies from 0.5 micron to 10 micron (1961-1985)
\ref De Vaucouleurs, G., De Vaucouleurs, A., Corwin JR., H.G., Buta, R. J. Paturel, G., 
 Fouque, P., 1991, Third reference catalogue of bright galaxies, version 3.9
\ref Diaz,A.I., Prieto, M.A., Wamsteker, W.  1988, A\&A,195, 53
\ref Dwek, E. 1987, ApJ, 322, 812
\ref  Egami, E.,  Neugebauer, B.,  Soifer, B.T.,  Matthews, K.,
  Becklin, E.E.,  Ressler, M.E.  2006, AJ,  131, 125
\ref Engel, H. et al. 2010, A\&A, 524 ,56). 
\ref Feldman U. 1992, Physica Scripta 46, 202
\ref Forster, N.M., Schreiber,  Roussel, H.  Sauvage, M.,   Charmandaris, V. 2004, A\&A, 419, 501 
\ref Fosbury, R.A.E., Wall, J.V. 1979, MNRAS, 189, 79
\ref Fried, J.W., Schulz, H. 1983, A\&A, 118, 166
\ref Glover, S.C.O., Mac Low, M.-M 2011, MNRAS, 412, 337
\ref Gallimore,J.F., Beswick, R. 2004, AJ, 127, 239
\ref Genzel, R. et al 1998, ApJ, 498, 579
\ref Golombek, D. , Miley, G.K.,   Neugebauer, G. 1988, AJ, 95, 26
\ref Gonzalez-Martin, O.; Masegosa, J.; Marquez, I.; Guainazzi, M.;
Jimenez-Bailon, E., 2009, A\&A, 506, 1107
\ref Gonzalez-Martin, O.; Masegosa, J.; Marquez, I.; Guerrero, M. A.;
Dultzin-Hacyan, D. 2006, A\&A, 460, 45
\ref  Gower, J.F.R., Scott, P.F., Wills, D.  1967, Mem. R. A. S.,  71,  49
\ref   Griffith, M.R., Wright, A.E., Burke, B.F., Ekers, R.D.  1995, ApJS, 97, 347
\ref Hagiwara, Y., Baan, W.A., Kl\"{o}ckner, H.-R. 2011, AJ, 142, 17
\ref Heckman, T.M., Armus, L., Miley, G.K. 1990, ApJS, 74, 833
\ref Joseph, R.D., Wright, G.S., Wade, R. 1984, Nature, 311, 132
\ref Klaas, U. et al. 2001, A\&A, 379, 823
\ref Komossa, S., Burwitz, V., Hasinger, G., Predehl, P., Kaastra, J.S., Ikebe, Y. 2003, ApJ, 582, L15
 \ref  Iono, D. et al.  2007, ApJ, 659, 283 
\ref  Laine S.,  Kotilainen, J.K., Reunanen, J., Ryder, S.D.,
 Beck, R. 2006, AJ, 131, 701 
\ref  Lahuis, F. et al. 2007, ApJ, 659, 296
\ref  Large, M.I., Mills, B.Y., Little, A.G., Crawford, D.F.,
Sutton, J.M. 1981,  MNRAS, 194, 693
\ref Luhman, M.L.  et al. 1998, ApJ, 504, L11 
\ref Lutz, D.; Maiolino R.; Spoon, H.; Moorwood, F. 2004, A\&A, 418, 465
\ref Mac Low, M.-M., Glover, S. C.O. 2012, MNRAS.tmp.2154
\ref Marshall, J. A.; Herter, T. L.; Armus, L.; Charmandaris, V.; Spoon, H. W.
W.; Bernard-Salas, J.; Houck, J. R. 2007, ApJ, 670, 129
\ref Moshir, M. et al 1990 vol. p.
\ref Nakanishi, K., Okumura, S. K., Kohno, K.,  Kawabe, R.,
Nakagawa, T. 2005, PASJ, 57, 575
\ref  Negishi, T.,  Onaka, T., Chan,  K.-W.,  Roellig, T.L.  2008, ApJS, 178, 280 
\ref Osterbrock, D.E. 1989 in Astrophysics of gaseous nebulae and active galactic nuclei. / University Science Books, 1989
\ref  Radovich, M., Ciroi, S., Contini, M., Rafanelli, P., Afanasiev, V. L., Dodonov, S. N.	
	 2005, A\&A, 431, 813	
\ref Rieke, G.H., Cutri, Roc M., Black, J.H., Kailey, W.F., McAlary, C.W. Lebofsky, M.J., Elston, R. 1985, 290, 116
\ref 	Rigopoulou, D., Spoon, H. W. W., Genzel, R., Lutz, D., Moorwood, A. F. M.,
 Tran, Q. D.   1999, AJ, 118, 2625	
\ref  Rodr\'{i}guez-Ardila, A.; Contini, M.; Viegas, S. M.  2005, MNRAS, 357, 220	
\ref Rudy, R. J., Levan, P. D.,  Rodriguez-Espinosa, J. M. 1982, AJ, 87,  598
\ref Sanders, D.B.,  Mazzarella, J.M.,  Kim, D. -C., Surace, J.A., Soifer, B.T.
2003, AJ, 126, 1607
\ref Sargsyan, L., Weedman, D., Lebouteiller, V., Houck, J., Barry, D., Hovhannisyan, A., Mickaelian, A. 2011, ApJ, 730, 19
\ref Sazonov, S.; Revnivtsev, M.; Krivonos, R.; Churazov, E.; Sunyaev, R. 2007, A\&A, 462, 57
\ref Siebenmorgen, R.,  Krugel, E., Spoon, H.W.W. 2004, A\&A, 414, 123 
\ref Spinoglio, L., Andreani, P., Malkan, M.A. 2002, ApJ, 572, 105S
\ref Spinoglio, L., Malkan, M.A., Rush, B., Carrasco, L., Recillas-Cruz, E. 1995, ApJ, 453, 616
\ref  Sugai, H.,  Malkan, M.A.,  Ward, M.J.,   Davies, R.I. and  McLean, I.S. 1997,
 ApJ, 481, 186
\ref Tacconi, L.J., Genzel, R., Tecza, M., Gallimore, J.F., Downes, D., Scoville, N.Z. 1999,  ApJ, 524, 732
\ref Tecza, M., Genzel, R., Tacconi, L.J. et al. 2000, ApJ, 537, 178
\ref Torrey, P., Cox, T.J.,  Kewley, L., Hernquist, L.  2012,  ApJ, 746,  108
\ref  Veilleux, S. et al.  2009, ApJS, 182, 628
\ref  Viegas, S. M., Contini, M., Contini, T.	 1999, A\&A, 347, 112	
\ref   Weaver, K. A. et al. 2010, ApJ, 716, 1151 
\ref   White,R.L., Becker, R.H. 1992, ApJS, 79, 331
\ref   Wilson, C. D. et al. 2008, ApJS, 178, 189
\ref   Wright, A., Otrupcek, R. 1990
    Parkes catalogue, 1990, Australia telescope national facility, PKSCAT90, 1990 
\ref Yang, M., Phillips, T. 2007, ApJ, 662, 284 

\appendix

\section{The models}
\begin{table}
\caption{The grid of selected models}
\begin{tabular}{cccccccc} \hline  \hline
\ [OII]/\Hb& [OIII]/\Hb & [OI]/\Hb & \Vs $^2$& \n0 $^3$ & $F$ $^4$ & $str$ & $D^5$\\ 
\  12.3$^1$ &   1.8$^1$ & 1.41$^1$ &   -       &    -      &  -   & -    & - \\ \hline
\ 11.      & 5.5        &   1.47   &  900      &  40        & 3.e8  &  0 & 1.3\\    
\ 4.6      & 2.56       &   0.27   &  700      &  40        & 3.e8  &  0 & 1.\\ 
\ 7.5      & 4.6        &   0.22   &  500      &  30        & 3.e8  &  0 & 1.\\
\ 3.76     & 2.11       &   0.2    &  500      &  40        & 3.e8  &  0 & 1. \\
\ 8.3      & 4.29       &   0.24   &  500      &  40        & 1.e8  &  0 & 1. \\
\ 19.3     & 9.48       &   0.4    &  500      &  40        & 5.e7  &  0 & 1.\\
\ 5.0      & 2.78       &   0.3    &  500      &  60        & 3.e8  &  0 & 1.\\
\ 7.       & 4.25       &   0.48   &  500      &  100       & 3.e8  &  0 & 0.5\\
\ 19.      & 11.        &   0.68   &  500      &  100       & 1.e8  &  0 & 0.3\\
\ 2.9      & 1.7        &   0.22   &  200      &  150       & 6.e8  &  0 & 0.5   \\
\ 6.7      & 4.         &   0.38   &  200      &  150       & 4.e8  &  0 & 0.33  \\
\ 8.       & 4.8        &   0.4    &  200      &  150       & 3.e8  &  0 & 0.5  \\
\ 4.       & 2.55       &   0.25   &  150      &  200       & 6.e8  &  0 & 0.4   \\
\ 17.      & 8.2        &   0.27   &  150      &  200       & 1.e8  &  0 & 0.5   \\ \hline
\ 12.      & 4.3        &   1.2    &  900      &   40       & 3.e8  &  1 & 0.6   \\ 
\ 10.6     & 4.7        &   1.34   &  900      &  60        & 3.e8  &  1 & 0.6   \\
\ 4.6      & 1.3        &   0.7    &  700      &   40       & 1.e9  &  1 & 1.   \\
\ 4.7      & 2.23       &   0.4    &  600      &  40        & 1.e8  &  1 & 1.   \\
\ 5.3      & 1.8        &   0.65   &  600      &  40        & 3.e8  &  1 & 1.   \\
\ 7.2      & 1.9        &   1.17   &  600      &  40        & 1.e9  &  1 & 1.   \\     
\ 7.       & 2.15       &   0.3    &  500      &  30        & 3.e8  &  1 & 1.   \\
\ 9.7      & 4.26       &   0.46   &  500      &  40        & 3.e7  &  1 & 1.   \\
\ 8.6      & 2.87       &   1.     &  500      &  40        & 2.e8  &  1 & 1.   \\
\ 11.      & 2.5        &   3.     &  500      &  40        & 3.e8  &  1 & 1.   \\  
\ 9.4      & 2.5        &  1.3     &  450      &  40        & 3.e8  &  1 & 1.   \\
\ 8.6      & 2.1        &   1.33   &  500      &  40        & 4.e8  &  1 & 1.   \\
\ 0.3      & 0.25       &   0.005  &  500      &  100       & 3.e7  &  1 & 1.   \\
\ 0.3      & 0.25       &   0.034  &  500      &  100       & 1.e8  &  1 & 1.   \\
\ 0.38     & 0.25       &   0.064  &  500      &  100       & 3.e8  &  1 & 1.   \\
\ 7.4      & 2.3        &   0.45   &  200      &  40        & 3.e8  &  1 & 1.   \\
\ 4.2      & 1.3        &   0.95   &  200      &  150       & 1.e8  &  1 & 1.   \\
\ 8.       & 1.2        &   2.6    &  200      &  150       & 6.e8  &  1 & 1.   \\
\ 4.6      & 1.26       &   0.48   &  150      &  200       & 1.e8  & 1  & 1.   \\ \hline
\ 35.      & 15.7       &   0.48   &  600      &  40        & 0.0   & SD & 1.   \\
\ 29.      & 16.        &   0.67   &  500      &  100       & 0.0   & SD & 1.   \\
\ 82.      & 27.9       &   0.52   &  200      &  40        & 0.0   & SD & 1.   \\
\ 90.      & 35.7       &   0.66   &  150      &  200       & 0.0   & SD & 1.   \\ \hline
\end{tabular}
 $^1$ observation data by FW79; $^2$ in \kms; $^3$ in \cm3; $^4$ in photons cm$^{-2}$ s$^{-1}$ eV$^{-1}$ at the Lyman limit; $^5$ in 10$^{17}$ cm

\end{table}

\end{document}